\documentclass[aps,prd,showkeys,showpacs,twocolumn,superscriptaddress]{revtex4}

\usepackage{graphics}
\usepackage{amsfonts}
\usepackage{amsmath}
\usepackage{latexsym}
\usepackage{amssymb}
\usepackage[mathcal]{euscript}
\usepackage{eurosym}
\usepackage{rotating}

\newcommand{\ket}[1]{| #1 \rangle}
\newcommand{\bra}[1]{\langle #1 |}
\newcommand{\braket}[2]{\langle #1 | #2 \rangle}
\newcommand{\trace}{\textrm{Tr}}

\newcommand{\europ}{\textup{\emph{\geneuro}}}

\newcommand{\whdoll}{\widehat{\$}}

\newcommand{\lonewh}{\widehat{\phantom{U}}}

\newtheorem{Def}{Definition}
\newtheorem{Iso}{Isomorphism}
\newtheorem{Th}{Theorem}
\newtheorem{Cl}{Claim}
\newtheorem{Lem}{Lemma}
\newtheorem{Proposition}{Proposition}
\newtheorem{Scenario}{Scenario}

\begin{document}

\title{\begin{center}
Quantum Decoys
\end{center}}

\author{Pablo Arrighi}
\email{pja35@cam.ac.uk} \affiliation{Computer Laboratory,
University of Cambridge, 15 JJ Thomson Avenue, Cambridge CB3 0FD,
U.K.}



\begin{abstract}
Alice communicates with words drawn uniformly amongst
$\{\ket{j}\}_{j=1..n}$, the canonical orthonormal basis. Sometimes
however Alice interleaves quantum decoys
$\{\frac{\ket{j}+i\ket{k}}{\sqrt{2}}\}$ between her messages. Such
pairwise superpositions of possible words cannot be distinguished
from the message words. Thus as malevolent Eve observes the
quantum channel, she runs the risk of damaging  the superpositions
(by causing a collapse). At the receiving end honest Bob, whom we
assume is warned of the quantum decoys' distribution, checks upon
their integrity with a measurement. The present work establishes,
in the case of individual attacks, the tradeoff between Eve's
information gain (her chances, if a message word was sent, of
guessing which) and the disturbance she induces (Bob's chances, if
a quantum decoy was sent, to detect tampering). Whilst not
directly applicable to secure channel protocols, quantum decoys
seem a powerful primitive for constructing other $n$-dimensional
quantum cryptographic applications. Moreover the methods employed
in this article should be of strong interest to anyone concerned
with the old but fundamental problem of how much information may
be gained about a system, versus how much this will disturb the
system, in quantum mechanics.
\end{abstract}

\maketitle
\section{Motivation and Claim}
The key principle of quantum cryptography could be summarized as
follows. \emph{Honest parties play using quantum states. To the
eavesdropper these states are random and non-orthogonal. In order
to gather information she must measure them, but this may cause
irreversible damage. Honest parties seek to detect her mischief by
checking whether certain quantum states are left intact.} The
tradeoff between the eavesdropper's information gain (about an
ensemble of quantum states), and the disturbance she necessarily
induces (upon this ensemble), can thus be viewed as
the power engine behind quantum cryptographic protocols.\\
Yet while numerous protocol-specific proofs of security have been
given, Information Gain versus Disturbance tradeoffs themselves
have remained stubbornly difficult to quantify. The problem was
first taken over by Fuchs and Peres\cite{Fuchs}, who tackled the
seemingly simple case of the two non-orthogonal equiprobable
states ensemble $\{(1/2, \ket{\psi_0}\bra{\psi_0}), (1/2,
\ket{\psi_1}\bra{\psi_1})\!\}$. A geometrical derivation of their
result can be found in \cite{cones}. For discrete distributions
this is just about the only result available. Of lesser interest
for cryptography, but very important in terms of its methods is
the work by Banaszek\cite{Banaszek}, who quantified the tradeoff
for the continuous uniform $n$-dimensional ensemble.
Barnum\cite{Barnum} makes several accurate qualitative remarks
upon the same ensemble, suggesting the tradeoff remains unchanged
for a uniform
distribution over mutually unbiased states. \\
In the present work we quantify the disturbance induced upon the
uniform ensemble of $n$-dimensional states $\{(1/n^2,\rho_{jk}\}$,
where $j$ and $k$ range from $1$ to $n$, and $\rho_{jk}$ stands
for the density matrix  of pairwise superpositions
$\frac{(\ket{j}+i\ket{k})(\bra{j}-i\bra{k})}{2}$ (note that when
$j=k$ this is simply the basis state $\ket{j}\bra{j}$). When
making use of non-orthogonal states this is no doubt a natural
distribution to consider, and thus an important building block for
$n$-dimensional cryptographic protocols. Its $\pi/2$ phase renders
this `pairing ensemble' undistinguishable from the canonical
ensemble $\{(1/n, \ket{j}\bra{j})\}$, for they both have density
matrix $\,Id/n$ (the maximally mixed state). This feature enables
the honest parties to hide the pairwise superpositions within
classical messages as means of securing those, i.e. to use the
superpositions as `quantum decoys'. In such situations the
eavesdropper seeks to gather information about the classical
messages, not the decoys. Therefore we quantify her information
gain with respect to the canonical ensemble $\{(1/n,
\ket{j}\bra{j})\}$, as suits the following scenario best:
\begin{Scenario}[Quantum decoys]
Consider a quantum channel for transmitting $n$-dimensional
systems having canonical orthonormal basis $\{\ket{j}\}$. Suppose
Alice's message words are drawn from the canonical ensemble
$\{(1/n, \ket{j}\bra{j})\}_{j=1\ldots n}$, whilst her
\emph{quantum decoys} are drawn from the pairing ensemble
$\{(1/n^2,\rho_{jk}\}_{j,k=1\ldots n}$, with
$\rho_{jk}=\frac{(\ket{j}+i\ket{k})(\bra{j}-i\bra{k})}{2}$. Alice
sends Bob, over the quantum channel, either a message word or a
decoy. Suppose that Bob, whenever a quantum decoy $\rho_{jk}$ gets
sent, measures
\begin{equation}\{P_{intact}= \big( \frac{\ket{j}+i\ket{k}} {\sqrt{2}}\big)
\big(\frac{\bra{j}-i\bra{k}}{\sqrt{2}}\big)\, ,\quad
P_{tamper}=\mathbb{I}-P_{intact}\}
\end{equation}
so as to check for tampering. Suppose Eve is eavesdropping the
quantum channel, and has an interest in determining Alice's
message words.
\end{Scenario}
Results on cryptographic protocols involving discrete
distributions in $n$-dimensional quantum systems (where $n$ is
left to vary), remain relatively scarce to this day
\cite{Acin},\cite{Bruss1}-\cite{Cerf},\cite{Liang} and tend to
focus on mutually unbiased states. We hope our main result will
prove a useful contribution to this difficult line of research:
\begin{Cl}[Statement of security]\label{claim} Referring to
Scenario $1$, suppose Eve performs an individual attack such that,
whenever a message word gets sent, she is able to identify which
with probability $G$ (mean estimation fidelity).\\
Then, whenever a quantum decoy gets sent, the probability $D$
(induced disturbance) of Bob detecting the tampering is bounded
below under the following tight inequality:
\begin{equation}\label{claim eq}
D\geq\frac{1}{2}-\frac{1}{2n}\Big(\sqrt{G}+\sqrt{(n-1)(1-G)}\Big)^2
\end{equation}
For optimal attacks $G$ varies from $\frac{1}{n}$ to $1$ as $D$
varies from $0$ to $\frac{1}{2}\!-\!\frac{1}{2n}$.
\end{Cl}
The reminder of this paper is dedicated to proving the above
statement. The method is just as important as the result, since it
seems applicable to several similar problems in quantum
cryptography.

In section \ref{method} we provide the necessary mathematical
results required to prove Claim \ref{claim}. We recall, in
particular, a key inequality regarding scalar products of vectors
(first obtained in \cite{Banaszek}), as well as some powerful
formulae arising from the state-operator correspondence (first
obtained in \cite{ArrPat}). Section \ref{prelim} exploits the
latter formula to express the probability of Bob \emph{not}
detecting the tampering (induced fidelity) as a linear functional
upon the positive matrix corresponding to Eve's attack. This
brings about crucial simplifications, finally placing us in a
position to apply the inequality. We do so in Section \ref{optim},
and prove our claim.

\section{Mathematical Methods}
\label{method}

\emph{Notations.} We denote by $M_d(\mathbb{C})$ the set of
$d\times d$ matrices of complex numbers, and by
$\textrm{Herm}_d^{+}(\mathbb{C})$ its subset of positive matrices,
also referred to as the (non-normalized) states of a
$d$-dimensional quantum system. In this section we let
$\{\ket{i}\}$ and $\{\ket{j}\}$ be orthonormal bases of
$\mathbb{C}^m$ and $\mathbb{C}^n$ respectively, which we will
refer to as canonical.\\

The following result is a minor generalization of some steps by
Banaszek \cite{Banaszek}.
\begin{Proposition}[Inequality]\label{ineq}
Consider a vector of complex numbers $v=(a_{jr})_{jr}$ together
with a function $j:\mathbb{N}\longrightarrow \mathbb{N}$. We then
have:
\begin{equation*}
f\leq\big(\sqrt{g}+\sqrt{(m-1)(n-g)} \big)^2
\end{equation*}
With
\begin{align*}
&g=\sum_r |a_{j_{(r)} r}|^2\\
&f=\sum_r |\!\sum_{j=0}^{m-1} a_{jr} |^2
\end{align*}
And subject to $||v||^2=n$.
\end{Proposition}
\emph{Proof.} Further let
\begin{align*}
v_j&=(a_{jr})_r\quad;\quad v_{j_{(r)}}=(a_{j_{(r)} r})_r\\
v_j'&=(a_{jr})_r\;\;\textrm{with r such that } j_{(r)}\neq j
\end{align*}
and notice that $g=||v_{j_{(r)}}||^2$, $f=\sum_{ij}v_i . v_j^*$.
The Cauchy-Schwartz inequality yields:
\begin{align}
v_i . v_j^* &\leq ||v_i||\,||v_j||\nonumber\\
f&\leq (\sum_{j=0}^{m-1} ||v_j||)^2\nonumber\\
f&\leq (\sqrt{g}+\sum_{j=0}^{m-1} ||v_j'||)^2\label{subineq1}
\end{align}
The quadratic/arithmetic mean inequality yields:
\begin{align}
\frac{1}{m-1} \sum_{j=0}^{m-1}||v_j'||&\leq\sqrt{\frac{1}{m-1}\sum_{j=0}^{m-1} ||v_j'||^2}\nonumber\\
&\leq \sqrt{\frac{n-g}{m-1}}\label{subineq2}
\end{align}
Combining Inequalities (\ref{subineq1}) and (\ref{subineq2})
yields the lemma. $\hfill\Box$

Next we remind the reader of an isomorphism from quantum states to
quantum operations, which in quantum information theory dates back
to the work of Jamiolkowski \cite{Jamio} and Choi \cite{Choi}. The
correspondence was subsequently reviewed and taken further in
\cite{ArrPat}, where Proposition \ref{totaltrace} appears. First
we relate vectors of $\mathbb{C}^m\otimes \mathbb{C}^n$ to
endomorphisms from $\mathbb{C}^n$ to $\mathbb{C}^m$.
\begin{Iso}
\label{isomorphism1} The following linear map
\begin{align*}
\hat{}\;:\mathbb{C}^m\otimes \mathbb{C}^n &\rightarrow
End(\mathbb{C}^n \rightarrow \mathbb{C}^m)\\
A&\mapsto\hat{A}\\
\sum_{ij} A_{ij}\ket{i}\ket{j}&\mapsto \sum_{ij}
A_{ij}\ket{i}\bra{j}
\end{align*}
where $i=1,\ldots,m$ and $j=1,\ldots,n$, is an isomorphism taking
$mn$ vectors $A$ into $m\times n$ matrices $\hat{A}$.
\end{Iso}
Second we relate elements of $M_{mn}(\mathbb{C})$ to linear maps
from $M_n(\mathbb{C})$ to $M_m(\mathbb{C})$.
\begin{Iso}
\label{isomorphism2} The following linear map:
\begin{align}
\lonewh\;:\mathbb{C}^{mn}\otimes
(\mathbb{C}^{mn})^{\dagger} &\longrightarrow{End(M_{n}(\mathbb{C})\rightarrow M_{m}(\mathbb{C}))}\nonumber\\
\$&\longmapsto{[\whdoll:\rho\mapsto\whdoll(\rho)]}\nonumber\\
\textrm{such that} \;\; A B^{\dagger}
&\longmapsto{[\rho\mapsto\hat{A}\rho\hat{B}^{\dagger}]}\quad\textrm{i.e.}
\nonumber\\
\sum_{ijkl}
A_{ij}B^*_{kl}\ket{i}\ket{j}\bra{k}\bra{l}&\longmapsto{[\rho\mapsto
\sum_{ijkl}
A_{ij}B^*_{kl}\ket{i}\bra{j}\rho\ket{l}\bra{k}\;]}\nonumber
\end{align}
where $i,k=1, \ldots,m$ and $j,l=1,\ldots, n$, is an isomorphism.
\end{Iso}

\begin{Def}
A linear map $\Omega : M_m(\mathbb{C}) \rightarrow
M_n(\mathbb{C})$ is Completely Positive-preserving if and only if
for all $r$ and for all $\rho$ in
$\textrm{Herm}_{mr}^+(\mathbb{C})$, $(\Omega\otimes Id_r)(\rho)$
belongs to $\textrm{Herm}_{nr}^+(\mathbb{C})$.
\end{Def}
Completely Positive-preserving linear maps from quantum states in
$\textrm{Herm}_{n}^+(\mathbb{C})$ to quantum states in
$\textrm{Herm}_{m}^+(\mathbb{C})$ are exactly those which are
physically allowable. They correspond, via Isomorphism
\ref{isomorphism2}, to quantum states in
$\textrm{Herm}_{mn}^+(\mathbb{C})$:
\begin{Th}\emph{\cite{Choi}}
\label{operations as states} The linear operation $\whdoll :
M_n(\mathbb{C}) \rightarrow M_m(\mathbb{C})$ is \emph{Completely
Positive-preserving} if and only if $\$$ belongs to
$\textrm{Herm}_{mn}^+(\mathbb{C})$.
\end{Th}
\begin{Def} A linear map $\whdoll: M_n(\mathbb{C})\rightarrow
M_m(\mathbb{C})$ is \emph{Trace-preserving} if and only if for all
$\rho$ in $M_n(\mathbb{C})$, $\trace(\whdoll(\rho))=\trace(\rho)$.
\end{Def}
Completely Positive-preserving linear maps having unit probability
of occurrence on every input quantum state are exactly those which
are Trace-preserving. They correspond, via Isomorphism
\ref{isomorphism2}, to quantum states in
$\textrm{Herm}_{mn}^+(\mathbb{C})$ verifying
\begin{equation}\label{trace preserving condition}
\trace_1(\$)=Id_n.
\end{equation}
\begin{Proposition}[State-operator formulae]\emph{\cite{ArrPat}} \label{totaltrace}
Let $\whdoll$ a linear map from $M_n(\mathbb{C})$ to
$M_m(\mathbb{C})$, $\sigma$, $\rho$ two elements of
$M_n(\mathbb{C})$, $\kappa$, $\tau$ two elements of
$M_m(\mathbb{C})$. Then we have:
\begin{equation*}
\kappa\whdoll(\rho \sigma)\tau= \trace_2\big( (\kappa\otimes
\rho^t)\$(\tau\otimes \sigma^t)\big)
\end{equation*}
where $\trace_2$ denotes the partial trace over the second system
$\mathbb{C}^n$  in $\mathbb{C}^m \otimes \mathbb{C}^n$. In
particular this implies that for all $\rho \in M_n(\mathbb{C})$
and $\kappa \in M_m(\mathbb{C})$,
\begin{equation*}
\trace\big(\kappa\whdoll(\rho)\big)= \trace\big((\kappa\otimes
\rho^t)\$\big).
\end{equation*}
\end{Proposition}
As with many quantum cryptographic problems our analysis will
require a careful optimization of the fidelity induced by a
quantum operation $\whdoll$. By means of the above formulae we
shall be able to write the induced fidelity as a linear functional
upon $\$$. This step is crucial to the next section (Lemma
\ref{fidelity as func}).
\section{Preliminary calculations}\label{prelim}
\subsection{Information Gain}\label{subsec information}
There exists several well-motivated methods with which to quantify
Eve's information gain. The one we shall adopt focuses on her
ability to make a guess after the measurement. Compared with
Shannon's mutual entropy this measure is advantageously close in
nature to the notion of disturbance.
\begin{Def}[Mean estimation fidelity]
The \emph{mean estimation fidelity} of a generalized measurement
$\{\hat{A}_r\}$ with guesses $\{\ket{\psi_r}\}$ w.r.t to an
ensemble $\{(p_i,\ket{\phi_i}\bra{\phi_i})\}$ is defined by:
\begin{align*}
G&=\sum_{r,i}p(r,i) |\braket{\phi_i}{\psi_r}|^2\\
&=\sum_{r,i}p_i\bra{\phi_i}\hat{A}^{\dagger}_r\hat{A}_r\ket{\phi_i}
\trace(\ket{\phi_i}\braket{\phi_i}{\psi_r}\bra{\psi_r})
\end{align*}
\end{Def}
The mean estimation fidelity is to be understood as the average
fidelity between the measurer's \emph{guess} knowing outcome $r$
occurred (the $\ket{\psi_r}$'s) and the $i^{th}$ state which was
indeed originally sent to him (the $\ket{\phi_i}$'s).\\
Notice it
is justified to consider that Eve's preferred attack is a
generalized measurement. In general she could perform a quantum
operation, which leaves her the possibility to regroup several
measurement outcomes into one likelier outcome. But there is no
information to be gained by ignoring the break-up of the likelier
measurement outcome. In fact this would simply force some of the
$\ket{\psi_r}$'s to be equal: the induced disturbance can only be
made worse.\\
In our scenario Eve gathers information about the canonical
ensemble $\{(1/n, \ket{j}\bra{j})\}_{j=1..n}$, for which one
obtains
\begin{equation*}
G=\frac{1}{n}\sum_r\trace\big(\bra{j}\hat{A}_r^{\dagger}\hat{A}_r\ket{j}\,\ket{j}\braket{j}{\psi_r}\bra{\psi_r}\big)
\end{equation*}
Clearly Eve's optimal guess knowing outcome $r$ occurred is
$\ket{j_{(r)}}$ such that
$\bra{j_{(r)}}\hat{A}_r^{\dagger}\hat{A}_r\ket{j_{(r)}}=\max_j
\bra{j}\hat{A}_r^{\dagger}\hat{A}_r\ket{j}$.\\
As a consequence
\begin{align*}
G&=\frac{1}{n}\sum_r
\bra{j_{(r)}}\hat{A}_r^{\dagger}\hat{A}_r\ket{j_{(r)}}\\
&=\frac{1}{n}\sum_r
\trace\big(\hat{A}_r\ket{j_{(r)}}\bra{j_{(r)}}\hat{A}_r^{\dagger}\big)\\
&=\frac{1}{n}\sum_r
\trace\big(Id\!\otimes\!\ket{j_{(r)}}\bra{j_{(r)}}{A}_r{A}_r^{\dagger}\big)
\end{align*}
where we applied Proposition \ref{totaltrace}. This yields:
\begin{Lem}[Estimation as a linear functional]\label{estimation as func}
Let $\whdoll\equiv\{\hat{A}_r\}$ be a generalized measurements
with best guess $\ket{j_{(r)}}$, and $\$\equiv\{A_r\}$ its
corresponding quantum state. \\
Further let
\begin{equation*}
\europ=\frac{1}{n} \sum_r
Id\!\otimes\!\ket{j_{(r)}}\bra{j_{(r)}}\!\otimes\!\ket{r}\bra{r}
\end{equation*}
With $j_{(r)}$ such that
$\bra{j_{(r)}}\hat{A}_r^{\dagger}\hat{A}_r\ket{j_{(r)}}=\max_j
\bra{j}\hat{A}_r^{\dagger}\hat{A}_r\ket{j}$.\\
Then the mean estimation fidelity of $\whdoll$ with respect to the
canonical ensemble is given by
\begin{equation}\label{estimation as func eq}
G=\sum_r\trace\big(\europ
\;(A_r\!\!\otimes\!\ket{r})(A_r\!\!\otimes\!\ket{r})^{\dagger}\big).
\end{equation}
\end{Lem}
\vspace{2pt}

\noindent As we have seen the generalized measurement is
equivalently described, using Isomorphism \ref{isomorphism1}, by
$\{A_r\}$, a set of non-zero non-normalized $n^2$-dimensional
vectors. Further consider the larger vector $v=(A_{ijr})_{ijr}$,
i.e. with $r$ itself an index of the complex components. The
trace-preserving condition upon the generalized measurement is
easily seen to imply that $||v||^2$ should be equal to $n$. From
Lemma \ref{estimation as func} it is clear that when seeking an
upper bound for $G$ under this fixed norm constraint, we may
assume $v$ to take the form $v=(A_{jjr})_{jjr}$, because of the
identity matrix on the first subsystem of $\europ$. As we shall
explain in subsection \ref{subsec disturbance} this can be done at
no cost for the mean induced fidelity. This way we reach the
following Lemma:
\begin{Lem}[Information] \label{spectral information} Consider a generalized measurement
$\{\hat{A}_r\}$, $\sum_r \hat{A}_r^{\dagger} \hat{A}_r=Id$,
$\hat{A}_r$ diagonal for all $r$, acting upon an $n$-dimensional
system. Then the mean estimation fidelity w.r.t the canonical
ensemble verifies
\begin{equation*}
G\leq\frac{1}{n}\,g
\end{equation*}
with $g=\sum_{r} |A_{j_{(r)} j_{(r)} r}|^2$ and $j_{(r)}$ such
that $|A_{j_{(r)}j_{(r)}r}|^2=\max_j |A_{jjr}|^2$.
\end{Lem}

\subsection{Disturbance}\label{subsec disturbance}
The notion of disturbance refers to Bob's chances of detecting
Eve's alteration of the state originally sent. For this purpose
Bob can, at best, project the received state upon the span of the
original state. Thus the disturbance verifies $D=1-F$, where $F$
is the induced fidelity.
\begin{Def}[Induced fidelity]
The \emph{fidelity induced} by a quantum operation $\whdoll$ upon
an ensemble $\{(p_i,\ket{\phi_i}\bra{\phi_i})\}$ is defined by:
\begin{equation*}
F=\sum_i p_i
\trace(\ket{\phi_i}\bra{\phi_i}\whdoll(\ket{\phi_i}\bra{\phi_i}))
\end{equation*}
\end{Def}
The induced fidelity is to be understood as the average fidelity
between the output of the quantum operation (the
$\whdoll(\ket{\phi_i}\bra{\phi_i})$'s) and its input (the
$\ket{\phi_i}\bra{\phi_i}$'s). A straightforward application of
Proposition \ref{totaltrace} yields: (with $^*$ denoting
componentwise complex conjugation as usual)
\begin{equation}\label{simple fidelity}
F=\sum_i p_i
\trace\big(\big(\ket{\phi_i}\bra{\phi_i}\!\otimes\!\ket{\phi^*_i}\bra{\phi^*_i}\big)\$\big)
\end{equation}
In our scenario Eve is tested on the pairing ensemble
$\{(1/n^2,\rho_{jk}\}_{j,k=1\ldots n}$, with
$\rho_{jk}=\frac{(\ket{j}+i\ket{k})(\bra{j}-i\bra{k})}{2}$, for
which one obtains:
\begin{align}
4\,\rho_{jk}\!\otimes\!\rho^*_{jk}&=\ket{jj}\bra{jj}+\ket{jj}\bra{kk}+i\ket{jj}\bra{jk}-i\ket{jj}\bra{kj}\nonumber\\
&+\ket{kk}\bra{jj}+\ket{kk}\bra{kk}+i\ket{kk}\bra{jk}-i\ket{kk}\bra{kj}\nonumber\\
&-i\ket{jk}\bra{jj}-i\ket{jk}\bra{kk}+\ket{jk}\bra{jk}-\ket{jk}\bra{kj}\nonumber\\
&+i\ket{kj}\bra{jj}+i\ket{kj}\bra{kk}-\ket{kj}\bra{jk}+\ket{kj}\bra{kj}\nonumber\\
\rho_{jk}\!\otimes\!\rho^*_{jk}&+\rho_{kj}\!\otimes\!\rho^*_{kj}=\frac{1}{2}\big(\ket{jj}+\ket{kk}\big)\big(\bra{jj}+\bra{kk}\big)\nonumber\\
&\quad\quad\quad\quad\;\;\;\,+\frac{1}{2}\big(\ket{jk}-\ket{kj}\big)\big(\bra{jk}-\bra{kj}\big)\nonumber\\
\sum_{jk}\rho_{jk}\!\otimes\!\rho^*_{jk}\!&=\label{summed decoys}\\
\frac{1}{4}\sum_{jk}\Big(\!\big(\ket{j&j}\!+\!\ket{kk}\big)\big(\bra{jj}\!+\!\bra{kk}\big)\!+\!\big(\ket{jk}\!-\!\ket{kj}\big)\big(\bra{jk}\!-\!\bra{kj}\big)\Big)\nonumber
\end{align}
We now proceed to express Equation (\ref{summed decoys}) in terms
of projectors. Regarding the subspace of repeated indices we
observe that:
\begin{align*}
\sum_{jk}\big(\ket{jj}\!+\!\ket{kk}\big)&\big(\bra{jj}\!+\!\bra{kk}\big)=2\sum_{jk}\big(\ket{jj}\bra{jj}+\ket{jj}\bra{kk}\big)\\
&=2n\!\!\sum_j\ket{jj}\bra{jj}+2\big(\sum_j\ket{jj}\big)\big(\sum_j\bra{jj}\big)
\end{align*}
As regards the subspace of non-repeated indices the vectors
$\ket{jk}-\ket{kj}$ are already orthogonal to each other, so long
as we maintain $j<k$. Combining our newly found spectral
decomposition with Equation (\ref{simple fidelity}) yields:
\begin{Lem}[Fidelity as a linear functional]\label{fidelity as func}
Let $\whdoll$ be a quantum operation, and $\$$ its corresponding
quantum state. \\
Further let
\begin{align*}
\pounds&=\frac{1}{2n}P_{rep}+\frac{1}{2n}P_{\beta}P_{rep}\\
&+\frac{1}{n^2}\sum_{j<k}\big(\frac{\ket{jk}-\ket{kj}}{\sqrt{2}}\big)\big(\frac{\bra{jk}-\bra{kj}}{\sqrt{2}}\big)P_{nonrep}\\
\textrm{With}\;\;\;&P_{rep}=\sum_j \ket{j}\bra{j}\otimes\ket{j}\bra{j}\quad P_{nonrep}=Id\!-\!P_{rep}\\
     &\ket{\beta}=\frac{1}{\sqrt{n}}\sum_j \ket{jj}\quad\,\textrm{and}\,\quad P_{\beta}=\ket{\beta}\bra{\beta}
\end{align*}
Then the fidelity induced by $\whdoll$ upon the pairing ensemble
is given by
\begin{equation}\label{fidelity as func eq}
F=\trace(\pounds\;\$).
\end{equation}
\end{Lem}
Using Theorem \ref{operations as states} $\$$ is positive and may
be thus be written $\$=\sum A_r A_r^{\dagger}$, with $\{A_r\}$ a
set of non-zero non-normalized $n^2$-dimensional vectors. Further
consider the larger vector $v=(A_{ijr})_{ijr}$. The
trace-preserving condition upon $\whdoll$ is easily seen to imply,
via by Equation (\ref{trace preserving condition}), that $||v||^2$
should be equal to $n$. From Lemma \ref{fidelity as func} it is
clear that, when seeking an upper bound for $F$ under this fixed
norm constraint and if $n\geq 2$, we may assume $v$ to lie in the
subspace of projector $P_{rep}\otimes Id$. In other words $v$
takes the form $v=(A_{jjr})_{jjr}$. As we have explained in
subsection \ref{subsec information} this can be done at no cost
for the mean estimation fidelity. We then have, using Lemma
\ref{fidelity as func} still:
\begin{equation*}
F=\frac{1}{2n}\sum_{ir}
|A_{jjr}|^2+\frac{1}{2n^2}\sum_{jkr}A_{jjr}A^*_{kkr}
\end{equation*}
This way we reach the following Lemma:
\begin{Lem}[Disturbance] \label{spectral disturbance}Consider a generalized measurement
$\{\hat{A}_r\}$, $\sum_r \hat{A}_r^{\dagger} \hat{A}_r=Id$,
$\hat{A}_r$ diagonal for all $r$, acting upon an n-dimensional
system. Then the disturbance induced upon the pairing ensemble
verifies
\begin{equation*}
D\geq\frac{1}{2}-\frac{1}{2n^2}f
\end{equation*}
with $f=\sum_r |\!\sum_{j=0}^{n-1} A_{jjr} |^2$.
\end{Lem}

\section{Optimization and Conclusion}\label{optim}
We are now set to prove Claim \ref{claim}. From Proposition
\ref{ineq} we immediately have
\begin{equation*}
\frac{1}{2}-\frac{1}{2n^2}f\geq\frac{1}{2}-\frac{1}{2n^2}\big(\sqrt{g}+\sqrt{(n-1)(n-g)}
\big)^2.
\end{equation*}
Applying Lemma \ref{spectral information} and \ref{spectral
disturbance} yields
\begin{equation*}
D\geq\frac{1}{2}-\frac{1}{2n^2}\big(\sqrt{nG}+\sqrt{n(n-1)(1-G)}
\big)^2
\end{equation*}
which in turn is nothing but Inequality (\ref{claim eq}). A plot
of the curve is shown in the Figure below.
\begin{figure}[b]
\begin{center}
\includegraphics[5.65cm,0.85cm][10cm,5.8cm]{plotdisturb.eps}
\end{center}
\caption{\textbf{Information Gain versus
Disturbance.}\\\centerline{\emph{In grey $n=4$, in black
$n=50$.}}}
\end{figure}
As was the case with the continuous uniform
ensemble\cite{Banaszek} the generalized measurement family
\begin{equation*}
\{\hat{A}_r\},\qquad \hat{A}_r
=\sqrt{G}\ket{r}\bra{r}+\sqrt{\frac{1-G}{n-1}}\big(Id-\ket{r}\bra{r}\big)
\end{equation*}
saturates the tradeoff for any fixed $G=1/n\,..\,1$. This may come
as no surprise since the corresponding $n^2$ vectors $A_r$ verify
\begin{equation*}
A_r=\lambda\ket{rr}+\mu\ket{\beta},\quad\lambda\!=\!\sqrt{G}\!-\!\sqrt{\frac{1-G}{n-1}}\quad\mu\!=\!\sqrt{\frac{1-G}{n-1}}.
\end{equation*}
In other words these unit vectors $\{A_r\}$ can be thought of as
superpositions of Eve's two extreme attacks: on the one hand
$\lambda=1$ yields the projective measurement $\{\ket{r}\bra{r}\}$
maximizing the mean estimation fidelity, whilst on the other hand
$\mu=1$ yields the `do nothing' measurement $\{Id\}$ minimizing
the disturbance. Viewed from the perspective of Lemma
\ref{fidelity as func} Eve, as she seeks to be more conservative,
increases her component in the subspace of $P_{\beta}$.

The generalized measurement family `measure $\{\ket{r}\bra{r}\}$
with probability $p$ else leave it alone' does not saturate the
tradeoff, but linear combinations of \emph{pure states
corresponding to measurement elements} do. Stated in this simple
manner, our result suggests the state-operator correspondence
method developed in this paper could establish itself as a very
natural procedure for deriving quantum cryptographic security
bounds in general.

Finally we wish to point out that cryptographic applications of
quantum decoys have recently been investigated. An asymmetric
variant of secure computation, whereby \emph{Alice gets Bob to
compute some well-known function $f$ upon her input $x$, but wants
to prevent Bob from learning anything about $x$, makes crucial use
of the artefact for its security  \cite{Bqc}}.

\section*{Acknowlegments}
P.J.A  would like to thank Anuj Dawar for proof-reading, EPSRC,
Marconi, the Cambridge European and Isaac Newton Trusts for
financial support.

\end{document}